\documentclass[12pt]{article}
\usepackage{graphicx}
\usepackage{graphics}
\input epsf

\def\beqn{\begin{eqnarray}}
\def\eeqn{\end{eqnarray}}
\relax
\def\ba{\begin{array}}
\def\ea{\end{array}}

\def\beq{\begin{equation}}
\def\eeq{\end{equation}}
\def\bea{\begin{array}}
\def\eea{\end{array}}

\def\to{\rightarrow}

\def\[{\left[}
\def\]{\right]}
\def\({\left(}
\def\){\right)}

\def\sm0{{\widetilde{m}_0}}

\def\U1em{{U(1)_{\rm em}}}
\def\to{\rightarrow}

\def\sq2{\sqrt{2}}

\def\End{\end{document}}

\def\sm{{\tilde{m}}} 

\begin{document}

\title{\Large \bf Perspectives of detecting\\ CKM-suppressed top quark decays at ILC}

\author{\large J.~L. Diaz-Cruz$^1$, C. Pagliarone$^2$\thanks{Corresponding author. 
E-mail: pagliarone@fnal.gov} \bigskip \\
{\it  $^1$~Benem\'erita Universidad Aut\'onoma de Puebla (BUAP), Puebla, Mexico} \\  
{\it $^2$~Universit\'a di Cassino \& INFN Pisa, Italy}}
\maketitle
\date{ }

{\large


\begin{abstract}
Top quark decays are of particular interest as a mean to
test  the standard model (SM) predictions, both for the
dominant ($t\to b+W$) and rare decays ($t\to q+W, cV, cVV,c\phi^0,bWZ$). 
As the latter are highly suppressed, they become an excellent window
to probe the predictions of theories beyond the SM. In particular, in this paper, 
we evaluate the  corrections from new physics to the CKM-suppressed 
SM top quark decay $t\to q+W$ ($q=d,s$), both within the effective lagrangian
approach and the MSSM and we discuss the perspectives to probe those predictions 
at the ILC.
\end{abstract}

\section{Introduction.}   
 After the discovery of the top quark at Fermilab Tevatron Collider~\cite{cdf_top,d0_top}, 
experimental attention has been turned on the examination of its  
production mechanisms and decay properties.  
Within the Standard Model (SM), the top quark production cross section is  
evaluated with an uncertainty of the order of $\sim 15\%$, while it is 
assumed to decay to a $W$ boson and a $b$ quark almost $100\%$ of the time. 
Due to its exceedingly heavy mass, the top quark is expected to 
be somehow related to new physics, and measuring its properties may then serve  
as a window for probing physics beyond the SM~\cite{Peccei}.  
It is also considered that the top quark may give some clue
to understand the mechanism of electroweak symmetry breaking.

\noindent
The interactions of quarks and leptons, with gauge bosons, seem to be correctly  
described by the $SU(3)_c\times SU(2)_L\times U(1)_Y$ gauge theory, as  
plenty of experimental data shows~\cite{revsm}.  
At tree-level, SM neutral interactions are diagonal,  
however, flavor changing neutral currents (FCNC) can arise  
at loop level. 
The fact that FCNC $B$-meson decays have been already detected, at 
rates consistent with the SM~\cite{PDG}, represents a great  
success for the model itself. 
However, SM predictions for top quark related processes   
are strongly suppressed, although corresponding   
experimental bounds are rather weak.   
At the coming LHC it is important to study rare top quark decays, because about
$10^7 \div 10^8$ top pairs will be produced per year. 
Thus, rare decays with branching rations ($BR$) of order $10^{-5}\div 10^{-6}$
may be detectable, depending on the signal.
The presence of any hint for new top quark physics at LHC~\cite{tt-rev}, 
would motivate further study to clarify the implications of those effects 
at the next generation of collider experiments~\cite{tt-LC}.  

\noindent

\section{A survey of top decays in the SM and beyond} 

Because of the structure of the SM, the $W$ boson coupling to
fermion pairs ($td_iW^\pm$), is proportional to the CKM element
$V_{td_i}$. Thus the decay $t\to b+W$ dominates its $BR$'s.
Radiative corrections to this mode have been evaluated in the
literature, both in the SM and some extensions, mainly within the
minimal SUSY extension of the SM (MSSM). In general, such corrections
are at most of order $10\%$, and therefore difficult to detect at
hadron colliders, but may be at the reach of the ILC.

\noindent
On the other hand, the top decay into the light quarks $t\to W +d(s)$
is suppressed, as they are proportional to $V_{td(s)}$. Furthermore, it is
unlikely that these modes could be detected at all at hadron
colliders. Probably for this reason, the SM corrections to this mode 
have not been studied. However, in extensions of the SM, it may be
possible to get a large enhancement that could even make it detectable
at the ILC, as will be shown in the next section.

\noindent
FCNC top quark decays, such as 
$t\to c \gamma$, $t\to c g$, $t\to c Z$ and $t\to c \phi$  
have been studied, for some time, in the context of both the SM
and  new physics \cite{bmele-rev}. 
In the SM, the branching ratio of FCNC top decays  
is extremely suppressed, as it is summarized in table 1.
The rare top quark decay $t\to c+\gamma$ was calculated first
in ref. \cite{ourtcg} in the SM and some extensions, 
the result implied a suppressed $BR$, less than about $10^{-10}$, which
was confirmed when subsequent analysis \cite{nexttcg} that included the correct
top mass value and gave $BR(t \to c+\gamma)= 5 \times 10^{-13}$. 
The decays  $t\to c+Z$ and  $t\to c+g$ were also calculated in refs.
\cite{nexttcg}. The resulting branching ratios turned out to be 
$BR(t\to c+Z)=1.3 \times 10^{-13}$ and $BR(t\to c+g)=5 \times10^{-11}$. 
None of them seem detectable at LHC nor at the ILC.

\noindent
The top-charm coupling with the SM Higgs  $\phi^0$ could also be
induced at one-loop level \cite{SM-tch}. 
The resulting  branching ratio is given by $BR(t\to c+\phi^0)=10^{-15}$ ,
which does not seem detectable neither.
The FCNC top decays involving a pair of vector bosons in the final
state, $t \to cVV$, can also be of interest \cite{ourtcvv}.
Although one could expect such modes to be even more suppressed than 
the ones with a single vector boson,
the appearance of an intermediate scalar resonance, as in the previous
case, could enhance the $BR$. Furthermore,  because of the large top quark 
mass, it also seems possible to allow the tree-level decay  $t\to b+WZ$, 
at least close to threshold. 

\noindent
Some typical results for the top decays in the SM are summarized in Table
1. This table also includes, for comparision, the results for top
branching ratios from models beyond the SM, in particular from the
THDM-III and SUSY, which will be discussed in what follows.

\begin{table}[t!] 
\begin{center}  
\begin{tabular}{| c| c| c| c| }  
\hline\hline  
{\bf BR\ } & SM & THDM-III & MSSM \\
\hline 
$\mathbf{BR(t \to sW)}$ & $2.2 \times 10^{-3}$ & $\sim 10^{-3}$ & 
$4\times 10^{-3}$ \\
\hline 
$\mathbf{BR(t \to c\phi^0)}$ & $10^{-13}-10^{-15}$ & $\sim 10^{-2}$ &  
$10^{-5}-10^{-4}$\\ 
\hline 
$\mathbf{BR(t \to c \gamma)}$ & $5\times 10^{-13}$ & $< 10^{-6}$ &  $<10^{-7}$ \\
\hline 
$\mathbf{BR(t \to c Z)}$ & $1.3 \times 10^{-13}$ &  $< 10^{-6}$ & $<10^{-7}$ \\
\hline
$\mathbf{BR(t \to c g)}$ & $5\times 10^{-11}$ & $< 10^{-6}$ & $< 10^{-5}$ \\
\hline
$\mathbf{BR(t \to c \gamma\gamma)}$ & $<10^{-16}$ & $\sim 10^{-4}$ &  $<10^{-8}$ \\
\hline
$\mathbf{BR(t \to c WW)}$ & $2\times 10^{-13}$  & $10^{-4}-10^{-3}$ & 
--  \\
\hline
$\mathbf{BR(t \to cZZ)}$ & -- & $10^{-5}-10^{-3}$ & -- \\
\hline
$\mathbf{BR(t \to bWZ)}$ & $2\times 10^{-6}$ & $\simeq 10^{-4}$ & -- \\
\hline  
\end{tabular}  
\end{center}  
\vspace{-0.25cm} 
\caption{ Branching ratios for some CKM-suppressed and FCNC top quark 
decays in the SM and beyond, for $m_t=173.5-178$ GeV. Decays into a pair of 
massive gauge bosons include  finite width effects of final state unstable 
particles.}  
\vspace{-0.2cm} 
\end{table} 

\section{Detection of top decays at the ILC}   

The International Linear Collider (ILC) is a proposed new electron-positron collider. 
Together, with the Large Hadron Collider at CERN (LHC), it would allow physicists to 
explore energy regions beyond the reach of today's accelerators. 
The nature of the ILC's electron-positron collisions would give it the capability 
to answer compelling questions that discoveries at the LHC will raise, from the 
identity of dark matter to the existence of extra dimensions.
In the ILC's design, two facing linear accelerators, each 20 kilometers long, hurl 
beams of electrons and positrons toward each other at energy that will be around 
$\sqrt s=$ $500$ GeV. The present phenomenological work have been performed 
assuming a center of mass energy of $\sqrt s=$ $500$ GeV, and an integrated luminosity, 
assuming two running experiments, taking data at the same time for 4 years
(plus the year 0) with a total integrated luminosity of $1$ ab$^{-1}$~\cite{ourwork}.
Heavy quarks, $b-$jets and $c-$jets, are tagged using their well known unambiguous properties such as their 
mass and their long lifetimes. 
Tag light-quark jets is much more difficult but anyhow this is needed in 
order to get meaningful measurement of the CKM matrix elements.
The tecnique used, in the present work,{ is the so called Large Flavour Tagging Method (LFTM)~\cite{ourwork}.
Particles with large fraction $x_p=$ $2p/E_{cm}$, of the momentum, carry information about 
the primary flavour. Then it is possible to define a class of function 
$\eta^ì_q(x^ì_p)$ that represent the  probability, for a quark of a flavour $q$, 
to develop into a jet in which $i$ is the particle having the largest $x_p=$ $2p/E_{cm}$.
Tagging Efficiencies are then extracted with almost no reliance on the hadronization model, 
using a sample of $Z_0$ evaluating single tag and double tag probabilities.
Hadronisation symmetries are introduced, in the equation systems, in order to simplify the
calculations:    $\eta^{\pi^\pm}_d= \eta^{\pi^\pm}_u$, $\eta^{K^\pm}_s= \eta^{K^0}_s$, 
$\eta^{e^\pm}_d= \eta^{e^\pm}_u$, $\eta^{\Lambda(\bar \Lambda)}_d= \eta^{\Lambda(\bar\Lambda)}_u$ 
and so on.

\noindent
We have been looking for dilepton top candidates decays (DIL), not including 
$\tau$ leptons, with two high-$p_T$ leptons, 
$M_{\ell^+\ell^-}$ outside the $Z_0$ mass window. Then, the jet-tagging requirements 
were: one tagged $b-$jet, vetoing the presence of an extra $b-$jet and also 
vetoing the presence of a tagged $c-$jet. The discrimantion of $s-$jets 
from other light-quark-jets have been achieved using the LFTM.

\noindent
A preliminary estimation, based on $1$ $ab^{-1}$, using the DIL signature, shows that sensitivity 
up to branching ratios of ~$10^{-3}$ may be reached in the channel $t \rightarrow s W $ where 
the Effective Lagrangian Approach predicts branching rations up to  $10^{-2}$, making possible to investigate 
for physics beyond the Standard Model. 
Further work, to increase sensitivity, adding $W+\,$jets analysis  and the $\tau$ leptons 
have to be done.


\section{Conclusions and perspectives} 

Rare decays of the top quark can be interesting 
probes of new physics.
$BR(t\to s+W) \simeq 1.5 \times 10^{-3}$ is reached in the SM.
In the minimal flavor violation scheme, one can get:
$BR(t\to s+W) \simeq  10^{-2}$.
In the MSSM, we can get an enhancement of order 50\%, which 
may help to make it detectable at ILC.
ILC, assuming a center of mass energy of $\sqrt s=$ $500$ GeV, and a total
integrated luminosity, for $4$ years, two experiments running, of $1$ ab$^{-1}$
will be able to reach sensitivity up to $10^{-3}$.

\section{Acknowledgments}
This work was supported by NSF and CONACYT-SNI (Mexico). 
We want to thank Laszlo Jenkovszky and the Organizing Committee for their warm, 
kind and nice hospitality.

  

}
\end{document}